\documentclass[aps,prl,reprint,superscriptaddress]{revtex4-1}

\usepackage{graphicx}
\usepackage{color}
\usepackage{amsmath,amssymb}
\usepackage{bm}
\usepackage{upgreek}
\usepackage{xspace}
\usepackage[colorlinks,urlcolor=blue,citecolor=blue,linkcolor=blue]{hyperref}

\newcommand{\uvect}[1]{\hat{\boldsymbol{#1}}\xspace}

\begin{document}

\title{Observation of a quantum phase from classical rotation of a single spin}

\author{A.~A.~Wood}
\affiliation{School of Physics, University of Melbourne, Victoria 3010, Australia}
\author{L. C. L. Hollenberg}
\affiliation{School of Physics, University of Melbourne, Victoria 3010, Australia}
\affiliation{Center for Quantum Computation and Communication Technology, University of Melbourne, Victoria 3010, Australia}
\author{R. E. Scholten}
\affiliation{School of Physics, University of Melbourne, Victoria 3010, Australia}
\author{A. M. Martin}
\email{martinam@unimelb.edu.au}
\affiliation{School of Physics, University of Melbourne, Victoria 3010, Australia}

\date{\today}

\begin{abstract}
The theory of angular momentum connects physical rotations and quantum spins together at a fundamental level. Physical rotation of a quantum system will therefore affect fundamental quantum operations, such as spin rotations in projective Hilbert space, but these effects are subtle and experimentally challenging to observe due to the fragility of quantum coherence. Here we report a measurement of a single-electron-spin phase shift arising directly from physical rotation, without transduction through magnetic fields or ancillary spins. This phase shift is observed by measuring the phase difference between a microwave driving field and a rotating two-level electron spin system, and can accumulate nonlinearly in time. We detect the nonlinear phase using spin-echo interferometry of a single nitrogen-vacancy qubit in a diamond rotating at 200,000\,rpm. Our measurements demonstrate the fundamental connections between spin, physical rotation and quantum phase, and will be applicable in schemes where the rotational degree of freedom of a quantum system is not fixed, such as spin-based rotation sensors and trapped nanoparticles containing spins.

\end{abstract}
\maketitle

Physical rotation is one of the most ubiquitous elements of classical physics. Everything from galaxies to individual molecules rotate on a myriad of timescales, with fundamental effects on the physical processes in each system. Quantum systems are also affected by physical rotation, but in most cases meaningful observation or exploitation of the effects is challenging due to the difficulties associated with controllably rotating an addressable quantum system at rates comparable to the coherence time of the system. Solid-state spin qubits such as the nitrogen-vacancy (NV) center in diamond~\cite{doherty_nitrogen-vacancy_2013, schirhagl_nitrogen-vacancy_2014} provide promising testbeds of how classical rotation affects quantum systems~\cite{wood_quantum_2018}, due to their long coherence times of up to several milliseconds~\cite{balasubramanian_ultralong_2009} and the robust nature of the host substrate. As the natural quantization axis of the NV is set by the host diamond crystal orientation, rotating the crystal rotates the qubit, and the effects of other phenomena such as magnetic fields can be examined independently~\cite{wood_magnetic_2017, wood_$t_2$-limited_2018}. In this work we observe a phase shift between a single NV qubit in a classically rotating diamond and an external microwave field used to drive quantum rotations. Depending on the angle between the instrinsic axis of the NV, the microwave field and the axis of rotation, this rotationally-induced phase can accumulate nonlinearly, and can therefore be detected in a spin-echo measurement of the NV electron spin.  

Detecting phase shifts arising from physical rotation is of significant interest to quantum sensing~\cite{zhang_inertial_2016,degen_quantum_2017}. Theoretical work~\cite{maclaurin_measurable_2012, ledbetter_gyroscopes_2012, song_nanoscale_2018} has proposed using the Berry phase~\cite{berry_quantal_1984} arising from physical rotation of a tilted NV qubit as a gyroscopic sensor, while other proposals envisage gyroscopes based on detecting rotational shifts of the Larmor precession frequency of a proximal nuclear spin~\cite{ajoy_stable_2012}. Experimentally, realizations of Berry phase in NV systems has been constrained to stationary systems. Yale \emph{et. al.} used optical transitions to drive the NV spin along closed paths on the Bloch sphere~\cite{yale_optical_2016}, while other work used an additional off-resonant microwave driving field varied along a circuit in the rotating frame~\cite{zhang_experimental_2016,arai_geometric_2018}, similar to the first observations of Berry's phase in solid-state qubits~\cite{leek_observation_2007} and NMR systems~\cite{suter_berrys_1987}. Recent experimental work has simulated rotation with phase-shifted microwave pulses, emulating rotation of the microwave field~\cite{jaskula_cross-sensor_2019}.  

The principle challenge associated with detecting the effect of rotation on an NV electron spin, from Berry's phase or otherwise, even in the case of rapidly rotating proof-of-principle experiments rests on discerning the small rotational phase shifts from the effects of magnetic fields and temperature variations. In our work, the phase shift from rotation can be made to accumulate non-linearly due to the effective rotation of the microwave field in the physically rotating frame of the NV qubit. We are thus able to use the decoupling properties of the spin-echo sequence to extend the interferometric interrogation time so that direct measurement of a rotationally-induced phase shift of the electron spin becomes possible in a noisy environment.   

\begin{figure*}[t]
	\centering
		\includegraphics{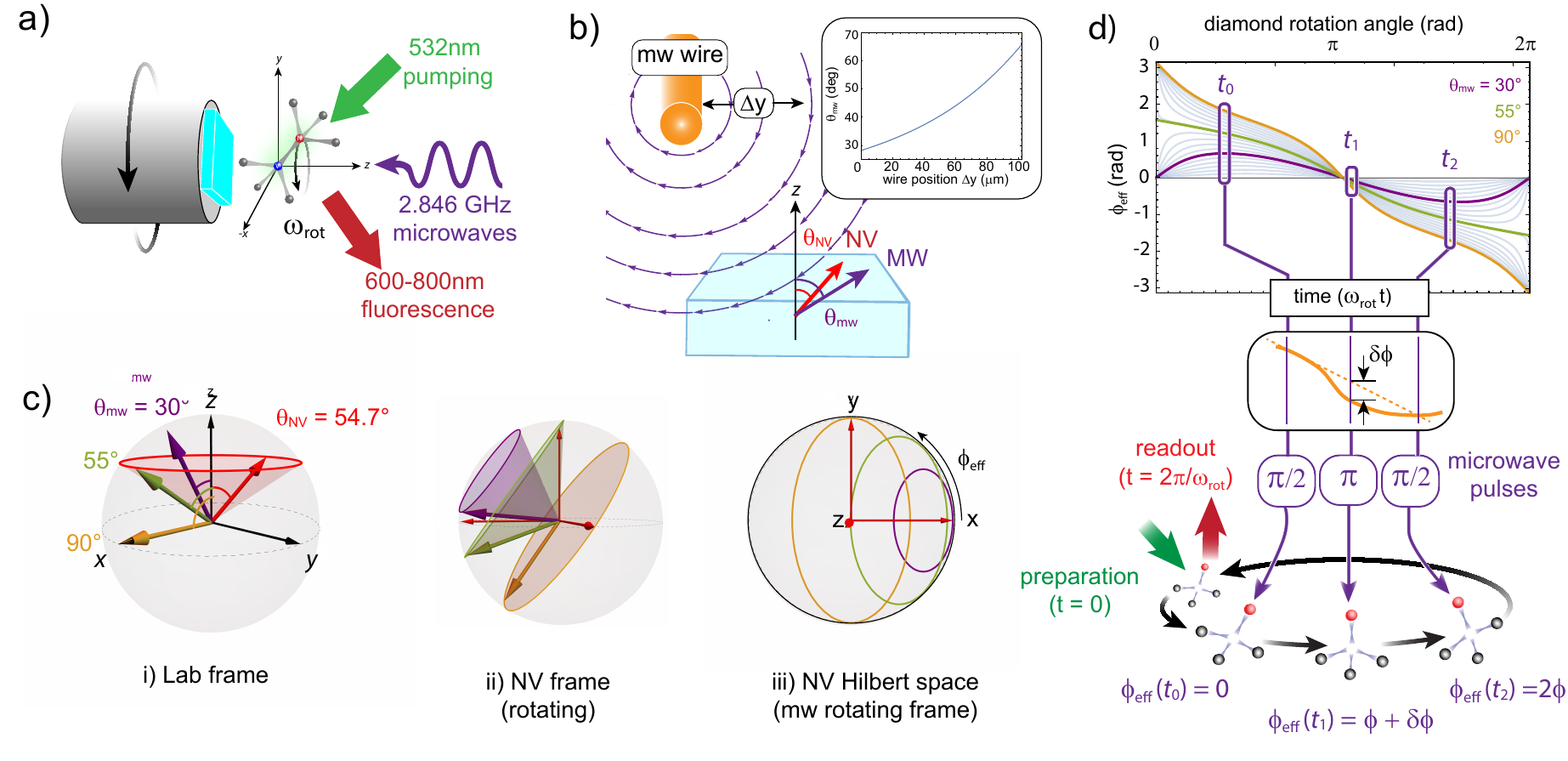}
	\caption{(a) Experimental setup, depicting NV center in a (100)-diamond mounted to an electric motor that rotates about an axis $z$ with azimuthal coordinate $\phi$. 532\,nm laser light is used to optically pump the NV spin, and microwaves resonant with the $m_S = 0\leftrightarrow m_S = -1$ transition are applied via a 20$\,\upmu$m diameter copper wire above the surface of the diamond. (b) The position of the microwave wire determines the angle the microwave polarization vector makes to the rotation axis, $\theta_\text{mw}$. Moving the microwave wire by up to $100\,\upmu$m along the $y$-axis varies $\theta_\text{mw}$ from $28^\circ$ to $67^\circ$. (c) The microwave vector at tilt angles of $\theta_\text{mw} = 30^\circ$ (purple), $54.7^\circ$ (green) and $90^\circ$ (orange) in the (i) stationary lab frame, (ii) NV frame, with $\theta_\text{NV} = 54.7^\circ$ and (iii) in the NV frame but rotating at the microwave frequency. The geometric path of the microwave vector, which reflects the physical rotation, is projected onto the azimuthal plane of the NV Bloch sphere under the rotating-wave approximation, where the azimuthal coordinate represents the overall effective phase $\phi_\text{eff}$ of the microwave field. (d) The effective phase $\phi_\text{eff}$ as a function of rotation angle $\phi$ (equivalently time $t$) for various $\theta_\text{NV}$, highlighting the nonlinear accumulation with rotation angle $\phi$. The phase of each pulse in a spin-echo sequence applied to the rotating qubit is set by $\phi_\text{eff}$, the linear phase $\phi$ is eliminated while the phase nonlinearity $\delta \phi$ introduced by rotation in the presence of the tilted microwave field is reflected in the spin-echo signal as $\cos^2\delta\phi$.}
	\label{fig:fig1}
\end{figure*}

 A schematic of our system is depicted in Fig.~\ref{fig:fig1} and the specific experimental configuration we use is similar to that described previously in Ref.\cite{wood_quantum_2018}. A $99.99\,\%$ $^{12}$C diamond is mounted on its $(100)$ face to an electric motor that spins at $\omega_\text{rot} = 3.33\,$kHz about an axis $z$. The diamond hosts individually resolvable NV centers with typical coherence times of $T_2^\ast \approx 50\,\upmu$s and $T_2 \approx 1\,$ms. Single NV centers are imaged by a scanning confocal microscope with illumination pulsed synchronously with the rotation of the motor. We consider a particular NV located about $3\,\upmu$m from the rotation center. The N-V axis is tilted by $\theta_\text{NV} = 54.7^\circ$ from $z$. Application of a magnetic field parallel to the rotation axis (not shown) breaks the degeneracy of the NV $m_S = \pm1$ states and allows us to consider the $|m_S=0\rangle$ and $|m_S = -1\rangle$ states as a pseudospin-1/2 system (see the Supplementary Information for full details). We introduce rotation operators $R_i(\theta) = \exp\left(-i S_i \theta\right)$ with $S_i = \frac{1}{2}\sigma_i$ the spin-1/2 Pauli matrices and $i\in\{x,y,z\}$, $\hbar = 1$. In the stationary laboratory frame, the rotating NV Hamiltonian is given by $R_\text{NV} H_0 R^{-1}_\text{NV}$ with $R_\text{NV} = R_z(\phi)R_y(\theta_\text{NV})$, $H_0 = D_\text{zfs} S_z^2$, $D_\text{zfs} = 2.870\,\text{GHz}$ the zero-field splitting and $\phi = \omega_\text{rot} t$ the azimuthal coordinate. The linearly-polarized microwave driving field can be represented as an oscillating magnetic field tilted from the $z$-axis by an angle $\theta_\text{mw}$, as depicted in Fig. \ref{fig:fig1}(c), and is represented by the operator $H_\text{mw}(\theta_\text{mw}) = R_y(\theta_\text{mw}) \Omega_0 S_z\,\cos\left(\omega t - \phi_0\right)R^{-1}_y(\theta_\text{mw})$, with $\Omega_0$ the Rabi frequency, $\omega$ the microwave oscillation frequency and $\phi_0$ an arbitrary initial phase. In the frame of the rotating NV, the interaction Hamiltonian is
\begin{equation}
H_I = R_\text{NV}^{-1}(\theta_\text{NV}, \phi) H_\text{mw}(\theta_\text{mw}) R_\text{NV}(\theta_\text{NV}, \phi).
\label{eq:hint}
\end{equation}
Under the rotating wave approximation (RWA), the diagonal terms of $H_I$ are neglected and only slowly-varying terms ($\ll\omega$) in the off-diagonal elements remain. A near-resonant microwave field is therefore approximated as an azimuthal vector in the rotating frame with components $\boldsymbol{\Omega} = (\Omega\,\cos\phi_\text{eff},\Omega\,\sin\phi_\text{eff},0)$, with $\Omega = |H^{(i,j)}_I|$ and $\phi_\text{eff} = \text{Arg}(H^{(i,j)}_I)$, \emph{i.e.} the modulus and argument of the complex numbers in the off-diagonal components of $H^{(i,j)}_I = \left(H^{(j,i)}_I\right)^\dagger$:
\begin{eqnarray}
H^{(i,j)}_I = & \Omega_0 e^{-i\phi_0}\left(\cos\theta_{\text{NV}}\cos\phi\sin\theta_{\text{mw}}\right. \nonumber \\ 
 - & \left. \cos\theta_\text{mw}\sin\theta_\text{NV}+ i\sin\theta_\text{mw}\sin\phi\right)/2.							
\label{eq:absarg}
\end{eqnarray}
In the frame rotating at the microwave frequency, the spin vector of the two-level system $S_i = \frac{1}{2}\langle \sigma_i\rangle$ precesses about the driving field $\boldsymbol{\Omega}$, and in a pulsed interferometric sequence it can be seen from Eq.(\ref{eq:absarg}) that the azimuthal rotation angle $\phi$ sets the effective phase $\phi_\text{eff}$ and determines the axis around which each pulse rotates the spin vector. For example, in a simple $\frac{\pi}{2}$-$\tau$-$\frac{\pi}{2}$ Ramsey sequence the relative phase of the second pulse of the sequence is therefore set by $\phi_\text{eff}$. 

The effective phase becomes more complicated when considering $\theta_\text{mw}\neq0$. The tilt angle of the microwave field $\theta_\text{mw}$ is varied by translating the position of a wire producing a microwave magnetic field, as shown in Fig.\ref{fig:fig1}(b). Figure \ref{fig:fig1}(c) depicts the tilted microwave field in the laboratory, rotating-NV frame and projective Hilbert space (the Bloch sphere, under the RWA), and the resulting effective phase $\phi_\text{eff}$ for different microwave tilt angles $\theta_\text{mw}$ when rotating. The geometric path followed by the microwave field in the NV frame (Fig \ref{fig:fig1}(c,ii)) is projected onto the azimuthal plane in the RWA Hilbert space (Fig \ref{fig:fig1}(c,iii)), with $\phi_\text{eff}$ identified as the azimuthal angle of $\boldsymbol{\Omega}$. For $\theta_\text{mw}\neq0$, $\phi_\text{eff}$ exhibits nonlinear behaviour, with $\boldsymbol{\Omega}$ oscillating in one hemisphere for $\theta_\text{mw} < \theta_\text{NV}$ before rotating about the entire azimuthal plane for $\theta_\text{mw} > \theta_\text{NV}$. Thus, for $\theta_\text{mw}\neq0$, $\phi_\text{eff}$ accumulates nonlinearly in rotation angle, allowing measurement with spin-echo interferometry. Spin-echo allows for longer interrogation times and therefore spans larger azimuthal angles, as well as insensitivity to temperature drifts compared to the simple Ramsey sequence.

A spin-echo sequence measures the difference in phase accumulation on either side of the refocusing $\pi$-pulse occuring midway between the initial and final $\pi/2$-pulses. Spin-echo is therefore insensitive to linearly accumulating phase shifts, whether originating in the quantum system itself (due to a static magnetic field, for instance) or the azimuthal precession of $\boldsymbol{\Omega}$ due to a linear $\phi_\text{eff}$. In the latter case, since the microwave field is switched off during the free evolution periods, it is the instantaneous $\phi_\text{eff}$ sampled by each microwave pulse that imparts the phase shift we seek to measure, and if $\phi_\text{eff}$ varies linearly across the pulse sequence, then it is cancelled in analogy with dc magnetic field shifts in conventional spin-echo measurements. Any deviation from linear effective phase accumulation (due to $\theta_\text{mw}\neq0$) when the $\pi$-refocusing pulse is applied (Figure \ref{fig:fig1}d) is reflected in the final populations of the two-level system, allowing us to define the nonlinear component of $\phi_\text{eff}$ as
\begin{equation}
\delta\phi  = \frac{\phi_\text{eff}(\tau)}{2} - \phi_\text{eff}\left(\frac{\tau}{2}\right),
\label{eq:pphase}
\end{equation}
assuming $\phi_\text{eff}(\tau = 0) = 0$. The population of the $|m_S = 0\rangle$ state then varies as $\cos^2\left(\delta\phi\right)$. Practically, we measure the rotationally-induced $\delta\phi$ by observing the phase of spin-echo interference fringes traced out as a function of some applied magnetic field.  In a rotating-NV spin-echo measurement, any dc magnetic field not parallel to the rotation axis is effectively up-converted to an oscillating field in the rotating frame~\cite{wood_$t_2$-limited_2018}. Since the effective microwave phase $\phi_\text{eff}$ can be controlled by varying the microwave tilt angle $\theta_\text{mw}$, the rotationally-induced phase shift manifests as a tilt-angle dependent fringe phase shift in a rotating-NV spin-echo measurement.  

\begin{figure}
	\centering
		\includegraphics{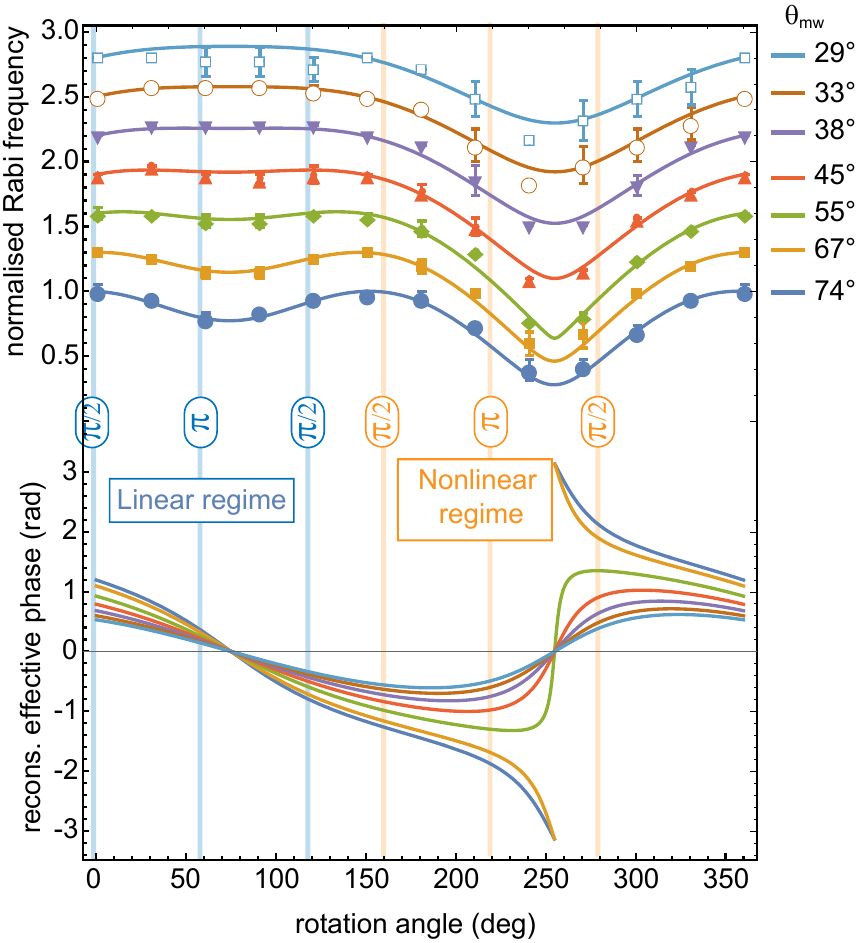}
	\caption{Reconstruction of effective phase accumulation from stationary Rabi frequency. Top: Measured normalized Rabi frequency vs. stationary rotation angle of motor for different microwave tilt angles, lines denote model derived from Eq.(\ref{eq:absarg}) and error bars denote uncertainty in calculated Rabi frequency derived from a Fourier transform of time-domain Rabi oscillations. Bottom: reconstructed effective phase, using the same model. Vertical lines denote the Rabi frequency and effective phase sampled where microwave pulses are applied -  for minimum (blue) and maximum (orange) nonlinearity of $\phi_\text{eff}$.}
	\label{fig:rabisphases}
\end{figure}

 As depicted in Fig.\ref{fig:fig1}(d), and according to Eq.(\ref{eq:pphase}), the specific behaviour of $\phi_\text{eff}$ sampled in a spin-echo measurement of duration $\tau$ may be approximately linear for a particular selection of microwave tilt angles, in which case $\delta\phi$ will be small, or the effective phase may be very nonlinear, resulting in a large $\delta\phi$. To estimate the expected $\phi_\text{eff}$ behaviour for our system we measured the angular dependence of the microwave coupling strength $\Omega(\theta_\text{mw}, \phi)$, which informs the vector behaviour of the microwave field and then reconstructed the expected phase behaviour. We measured the microwave Rabi frequency by performing a discrete Fourier transform of time-domain Rabi oscillations collected at different stationary `park' angles $\phi$, for various microwave wire positions ($\theta_\text{mw}$). Using a simple approximation of the microwave field emanating from the wire as $\uvect{B}_\text{mw} = \uvect{\varphi}$, where $\varphi$ is the azimuthal coordinate in the cylindrical coordinate system $(\uvect{r'}, \uvect{\varphi}, \uvect{x})$ of the wire, is sufficient to accurately reproduce the angular dependence of the measured Rabi frequency $\Omega$. The only free parameter in the plotted model depicted in Fig \ref{fig:rabisphases} is the calibration between absolute azimuthal orientation of the NV axis and the arbitrary motor park angle. These measurements also determine the microwave pulse durations required to execute the $\pi/2$- and $\pi$-pulses used in the rotating spin-echo sequence, which can differ significantly in a single sequence for certain $\theta_\text{mw}$. The maximum variation of Rabi frequency across a single spin-echo sequence was between $1$ and $6\,$MHz. Figure \ref{fig:rabisphases} shows the measured microwave Rabi frequency and reconstructed effective phase as a function of motor park angle. 

For our measurements of $\phi_\text{eff}$ while rotating, we chose $\tau =100\,\upmu$s ($120^\circ$ at $3.33\,$kHz) and two initial configurations: a `null' measurement where $\phi_\text{eff}$ is approximately linearly varying ($\delta_\phi \approx 0$), and a second configuration sampling the highly nonlinear behaviour in the vicinity of $\phi = 250^\circ$ to observe maximum $\delta\phi$. The experimental sequence was synchronized to the rotation of the diamond, and a delay time between the motor trigger and the start of the experimental sequence was used to change the starting angle in the spin-echo experiment. The experimental sequence consisted of measuring spin-echo fringes as a function of an applied dc magnetic field ${\bf B} = (B_x,0,0)$ for different microwave tilt angles $\theta_\text{mw}$. One preparation-interrogation-readout cycle was performed every rotation period, and was repeated $>10^5$ times over an approximate duration of two to three minutes.  

\begin{figure}
	\centering
		\includegraphics[width = \columnwidth]{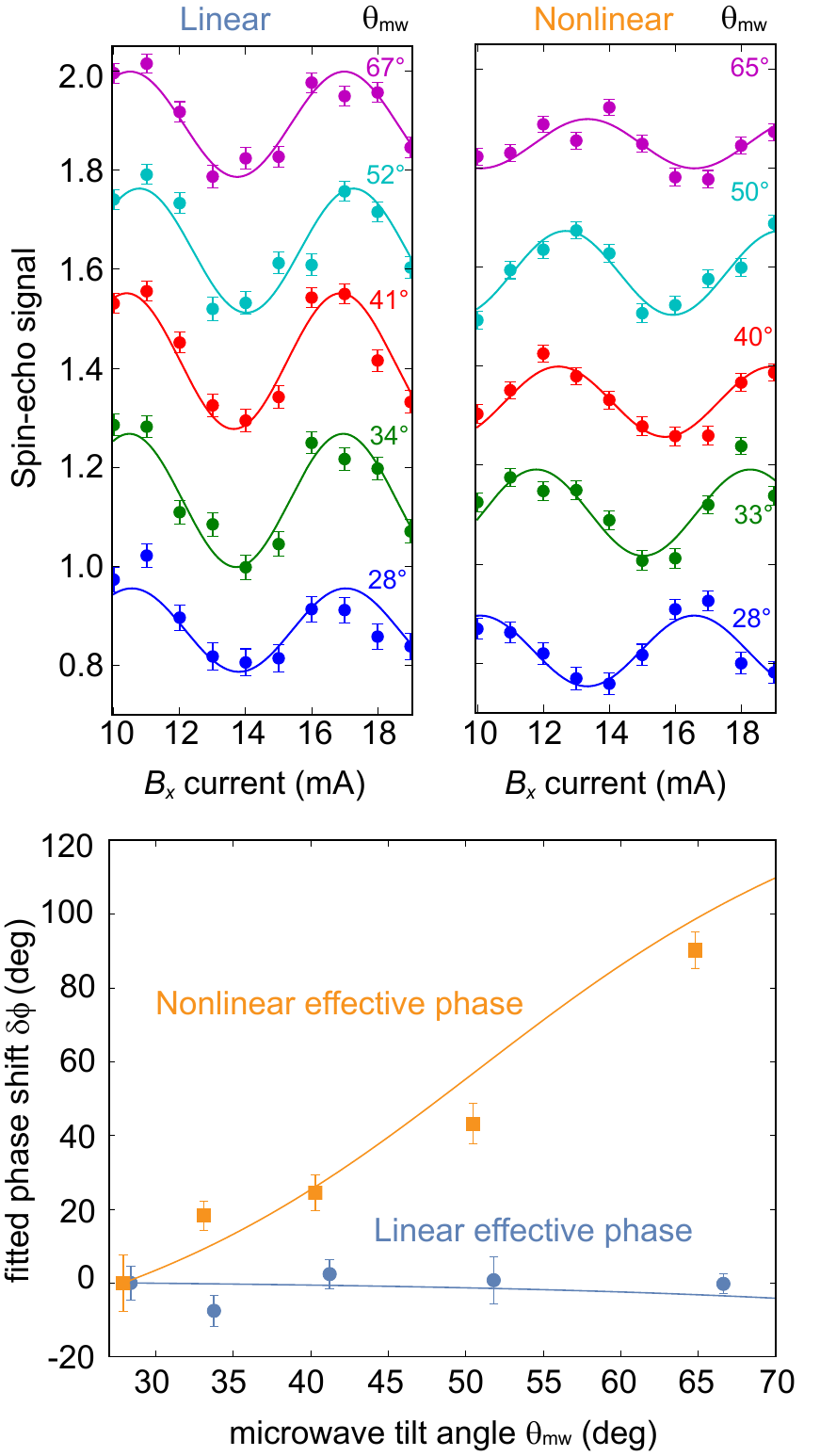}
	\caption{Detection of electron-spin phase shifts from physical rotation. Top: Spin-echo fringes with $\tau = 100\,\upmu$s ($\Delta\phi = 120^\circ$) as a function of applied $x$-field for different microwave tilt angles, for the linear measurement ($\phi_0 = 357^\circ$, left) and for a measurement maximally sensitive to the nonlinear part of $\phi_\text{eff}$ ($\phi_0 = 160^\circ$, right). Spin-echo signal error bars denote uncertainty in measured spin state, which is dominated by photon counting statistics. Bottom: fitted fringe phase shift vs. microwave tilt angle for linear (blue) and nonlinear (orange) regions. Error bars denote uncertainty in fitted phase.}
	\label{fig:fig3}
\end{figure}

Figure \ref{fig:fig3}(a,b) shows spin-echo fringes for several different microwave tilt angles in the regions of linear and nonlinear $\phi_\text{eff}$ accumulation. We extract the phase shift, plotted in Fig \ref{fig:fig3}(c), by fitting functions of the form $\cos^2(2\pi f_0-\delta\phi)$ to the fringe data, where $f_0$ is the average fringe frequency across the whole dataset. Drifts of the apparatus temperature, which in turn lead to drifts in the background magnetic field environment, are particularly problematic, since a drift in the transverse magnetic fields will result in a spurious shift of the spin-echo signal phase. For this reason, we sample the spin echo signal over a narrow range of $B_x$ (one period) and as rapidly as possible. A more detailed analysis of drifts and other sources of error is provided in the Supplementary Material. The results agree very well with the model developed from the stationary data in Fig \ref{fig:rabisphases}, and constitute the first demonstration of a single, room-temperature spin system with a quantum phase determined by classical rotation.   

In measuring a rotationally-induced phase, we have performed a detection of actual physical rotation using the electron spin of the NV center. The nonlinear aspect of $\phi_\text{eff}$, key to our measurement working in a noisy environment, is a somewhat overlooked effect that may offer new possibilities for precision rotation sensing or an additional control handle for quantum systems. While practical deployment of a diamond-based rotation sensor is a considerable technological challenge, our results will be applicable to other systems where physical rotation on quantum timescales plays an important role. On a much slower timescale than the work reported here, studies of Brownian motion and rotational diffusion of nanodiamonds in fluidic environments~\cite{mcguinness_quantum_2011,maclaurin_nanoscale_2013,yoshinari_observing_2013} offers an intriguing example of multi-axis physical rotations in quantum systems, and our work offers the prospect of taking such measurements a step further and probing rapid rotation and motion on quantum-relevant timescales. In a more general sense, our results serve to further highlight the remarkable physical effects that simple physical rotation can impart to sophisticated systems. Further work could consider quantum measurements of rotation on faster timescales in different systems, such as optically and electrically-trapped nanodiamonds~\cite{horowitz_electron_2012,hoang_torsional_2016, delord_electron_2017, delord_ramsey_2018} which have interesting prospects for realising sensitive torque detectors and probing fundamental quantum mechanics~\cite{delord_strong_2017, stickler_probing_2018, delord_spin-cooling_2019}.

\section*{Acknowledgements}
We thank L. P. McGuinness for provision of the diamond sample and L. D. Turner and R. P. Anderson for fruitful discussion. This work was supported by the Australian Research Council Discovery Scheme (DP150101704, DP190100949). 

\end{document}